\documentclass[useAMS]{mn2e}
\usepackage{amsmath,amssymb,epsfig,natbib_vw,times}

\bibpunct[, ]{(}{)}{;}{a}{}{,}

\def \aj {AJ}
\def \mnras {MNRAS}
\def \apj {ApJ}
\def \apjs {ApJS}
\def \apjl {ApJL}
\def \aap {A\&A}
\def \nat {Nature}
\def \araa {ARAA}
\def \pasp {PASP}

\def \Ledd {L$_{\rm Edd}$}
\def \ledd {L$_{\rm Edd}$}
\def \Msol {M$_\odot$}

\def \hii {H~{\sc ii}}

\def \ha {H$\alpha$}
\def \nii {[N~{\sc ii}]}
\def \oiii {[O~{\sc iii}]}
\def \loiii {L$_{\rm OIII}$}

\def \hb {H$\beta$}

\title[Timing the starburst-AGN connection]{Timing the starburst-AGN connection}
\author[V. Wild, T. Heckman, S. Charlot]{
\parbox[t]{\textwidth}{\raggedright 
Vivienne Wild$^1$\thanks{wild@iap.fr}, Timothy Heckman$^2$, St\'{e}phane Charlot$^1$
}
\vspace*{6pt}\\
$^1$ Institut d'Astrophysique de Paris, CNRS, Universit\'{e}
Pierre \& Marie Curie, UMR 7095, 98bis bd Arago, 75014 Paris, France\\
$^2$  Department of Physics and Astronomy, Johns Hopkins
University, Baltimore, MD 21218, USA  
}

\begin{document}

\maketitle
\begin{abstract}
  The mass of super massive black holes at the center of galaxies is
  tightly correlated with the mass of the galaxy bulges which host
  them.  This observed correlation implies a mechanism of joint
  growth, but the precise physical processes responsible are a matter
  of some debate.  Here we report on the growth of black holes in 400
  local galactic bulges which have experienced a strong burst of star
  formation in the past 600\,Myr.  The black holes in our sample have
  typical masses of $10^{6.5}-10^{7.5}$M$_\odot$ and the active nuclei
  have bolometric luminosities of order $10^{42}-10^{44}$erg/s.  We
  combine stellar continuum indices with \ha\ luminosities to measure
  a decay timescale of $\sim$300\,Myr for the decline in star
  formation after a starburst. During the first 600\,Myr after a
  starburst, the black holes in our sample increase their mass by
  on-average $5\%$ and the total mass of stars formed is about $10^3$
  times the total mass accreted onto the black hole. This ratio is
  similar to the ratio of stellar to black hole mass observed in
  present-day bulges. We find that the average rate of accretion of
  matter onto the black hole rises steeply roughly 250\,Myr after the
  onset of the starburst. We show that our results are consistent with
  a simple model in which 0.5\% of the mass lost by intermediate mass
  stars in the bulge is accreted by the black hole, but with a
  suppression in the efficiency of black hole growth at early times
  plausibly caused by supernova feedback, which is stronger at earlier
  times. We suggest this picture may be more generally applicable to
  black hole growth, and could help explain the strong correlation
  between bulge and black hole mass.

\end{abstract}


\section{Introduction}

The existence of a strong correlation between the masses of
supermassive black holes and the galactic bulges where they reside is
strongly suggestive of a mechanism of joint growth
\citep{2000ApJ...539L..13G, 2000ApJ...539L...9F,
  2002ApJ...574..740T}. However, the physical mechanism(s) responsible
for this observed relation remain unclear
\citep{Ferrarese:2005p2397,Cattaneo:2009p2507}. The very different
physical scales involved in building bulges through star formation and
feeding a central black hole pose a great challenge for the possible
mechanisms of causal connection. Several theoretical models
successfully reproduce the observed relation,
\citep[e.g.][]{Silk:1998p2398, Haehnelt:1998p2406,
  2006ApJS..163....1H, Ciotti:2007p1476,2008MNRAS.387...13K}, most
invoking some form of ``feedback'' by which the energy radiated during
the phase of rapid accretion onto the black hole prevents further
accretion and shuts down star formation. However, observationally
there is little evidence for effective direct ``mechanical'' feedback
caused exclusively by energy released during accretion onto the black
hole.

Clues to the nature of the relationship between the growth of bulges
and black holes have been provided over the last decade through the
significant observational progress that has been made in understanding
the demographics of the AGN population and its relationship to the
properties of the host galaxies. In the local Universe, the Sloan
Digital Sky Survey (SDSS) spectroscopic survey
\citep{2000AJ....120.1579Y, 2002AJ....124.1810S,Abazajian:2009p2485}
has provided the statistics necessary to calculate volume averaged
quantities and to reveal which populations of AGN and galaxies are the
sites of the majority of present day black hole
growth. \citet{2003MNRAS.346.1055K} showed that significant growth of
black holes is presently happening in galaxies with the masses and
structural properties of early-type galaxies, but with significant
amounts of recent or on-going star-formation.  \citet{heckman04} then
found that the majority of low-redshift black hole growth occurs in
high growth rate phases (L/\ledd\ greater than a few percent) which
must be of short duration relative to the Hubble time. This is
consistent with the fact that integrating the luminous output of
quasars (high L/\ledd\ accretors) over all cosmic times gives a black
hole mass density similar to that observed in the local Universe
\citep{Yu:2002p3731}. Comparing the integrated ongoing starformation
to the integrated ongoing black hole accretion, \citet{heckman04}
found a value close to that expected from the ${\rm M_{BH}-M_{bulge}}$
relation, implying that bulge formation and BH growth are still
tightly coupled even today. However, rapid growth of the population of
bulges and their black holes is occurring today only in systems of
relatively low-mass (so-called cosmic downsizing).

One long-standing idea linking the evolution of bulges and black holes
is that there is a strong evolutionary connection between intense
central bursts of star formation and the growth of black holes. In
\citet{wild_psb} we found that in fact the majority of the most
rapidly growing black holes at the present day exist in star-forming
galaxies with bulges, but with relatively ordinary recent star formation
histories and no indication of a recent strong burst of star
formation. However, we also found that galaxies which are undergoing
or have recently undergone the strongest bursts of star formation in
the sample showed an increased probability of hosting a rapidly
growing black hole.  This link between strong starburst and black hole
growth may have been more important at higher redshifts where gas was
more abundant (potentially leading to larger and more frequent bursts
of star formation), and where the majority of the black hole mass in
the universe was accreted.

Progress on understanding the link between star formation and black
hole growth has also been provided by detailed observations of very
nearby AGN which can probe spatial scales of 10s to 100s of parsecs
(an order of magnitude smaller than the typical scale probed by the
3'' SDSS fibre aperture). Evidence for young nuclear stellar
populations is found in around half of local AGN studied
\citep[see][for a recent summary of the
literature]{Davies:2007p1422}. In the future such an approach may be
extended to provide a systematic investigation covering several orders
of magnitude in bolometric luminosity and black hole mass, and
including starbursts with no current black hole growth.  Of direct
importance to this paper is the discovery of a possible delay between
the peak of the starburst and AGN activity for a handful of nearby AGN
\citep{Tadhunter:1996p3737,Emonts:2006p1603, Goto:2006p1655,
  Davies:2007p1422}.  Such a delay may also be related to the very low
accretion rate observed in the Galactic center
\citep{Baganoff:2003p3615}, despite the presence of considerable wind
material from young OB stars \citep{Paumard:2006p3638}.
  \citet{Schawinski:2007} used the Lick indices and photometry of a
  morphologically selected sample of elliptical galaxies to reveal
  that those with emission lines dominated by star formation typically
  have more recent star formation than those with emission lines
  dominated by AGN.  They interpreted this as a delay between
  starformation and AGN activity. However, they neglect the fact that
  more recent star formation will hide increasingly powerful AGN,
  which will significantly affect their result.

In this paper we bridge the gap between the gross trends of stellar
populations with AGN properties uncovered by SDSS, and detailed
studies of the recent star formation in the nuclei of local AGN. We
use the SDSS to select the galaxies with bulges which have recently
undergone the strongest bursts of starformation in the local
Universe. Our sample is complete for ages up to 600\,Myr after the
onset of the starburst, considerably longer than the typical timescale
for black hole accretion events. From this sample, we construct a
time-averaged view of black hole accretion subsequent to a star
formation episode.

In Section \ref{sec:2} we present the sample and the method used to
measure the age of the starburst from the stellar continuum. In
Section \ref{sec:3} we derive black hole accretion rates and
instantaneous star formation rates from the emission line fluxes, and
quantify the completeness of our samples. The results are presented in
Section \ref{sec:results} and discussed in Section
\ref{sec:disc}. Throughout the paper, we calculate luminosities using
$\Omega_{\rm M}=0.3$, $\Omega_\Lambda=0.7$ and ${\rm
  H_0}=70$km/s/Mpc. For the reader who is less interested in the
technical aspects of this work we recommend reading the introductions
to Sections 2 and 3 (i.e. skipping the subsections), before moving
onto the results and conclusions.

\section{A local starburst sample}\label{sec:2}

\begin{figure}
\includegraphics[width=89mm]{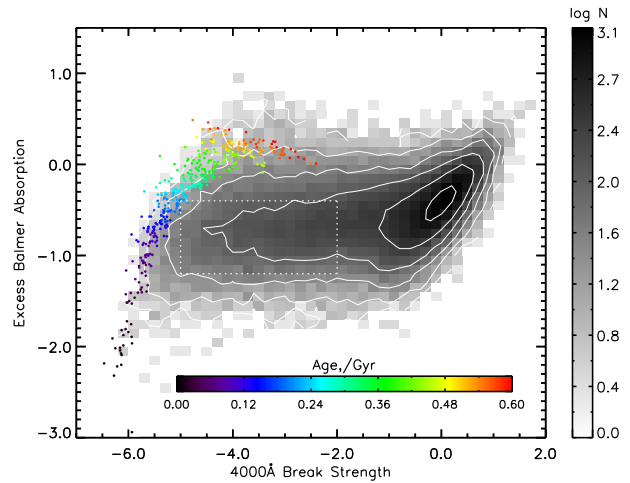}
\caption{The distribution of 4000\AA\ break strength and Balmer
  absorption line strength for 70,000 bulge-galaxies (grayscale, in
  logarithmic number). The spectral indices are generated using a
  Principal Component Analysis of a model spectral library as
  described in Wild et al. (2007). The star formation histories of
  galaxies that lie in different regions of the diagram are described
  in Section \ref{sec:2}. The starburst sample is overplotted as
  coloured dots, where colour indicates the age of the starburst.  The
  dotted box indicates the region from which the starforming
  comparison sample is drawn.}\label{fig:pca}
\end{figure}

AGN (the observational manifestation of growing black holes) are
found in galaxies with bulges, from early type spirals through to
massive ellipticals (Ho et~al. 1997; Dunlop et al. 2003; Kauffmann et
al. 2003; but see Filippenko \& Ho 2003 for an exception)
\nocite{Ho:1997p2415, 2003MNRAS.340.1095D, 2003MNRAS.346.1055K,
  Filippenko:2003p2451}. Therefore we select as our starting point a
base sample of $\sim$70,000 galaxies from the Sloan Digital Sky Survey
(SDSS) data release 7 \citep{Abazajian:2009p2485}, selected to have
high stellar surface mass densities characteristic of bulges
(Kauffmann et al. 2003).  We restrict the sample to galaxies with
redshifts $0.01<z<0.07$. At these redshifts the 3 arcsec aperture of
the SDSS optical fibres samples the inner $\lesssim$2kpc radius of the
galaxies, so that light from the stars in the bulge dominates the
spectrum \citep{Gadotti:2009p2216}. Broadline (i.e. Type 1 or
unobscured) AGN are not included in our sample, as light from the
central nucleus would contaminate our study of the stellar population
in the host galaxy. Full details of the sample selection criteria are
given in the following subsection, in the remainder of this section we
introduce the concept of our methodology.

In the integrated spectrum of a galaxy the different signatures of
stars of different ages can be used to obtain information about a
galaxy's recent star formation history.  In Fig. \ref{fig:pca} we
begin by inspecting the distribution of two common stellar continuum
indices for the 70,000 bulge-galaxies, the 4000\AA\ break strength and
the excess Balmer absorption line strength \citep{wild_psb}. The
majority of bulge-galaxies show no evidence of recent or current star
formation, they form the ``red-sequence'' which lies on the right,
with strong 4000\AA\ break strength indicating that the galaxies
contain predominantly old stars. There is a significant minority of
bulge-galaxies that are forming stars, these have younger mean stellar
ages and therefore weaker 4000\AA\ break strengths and form the
``blue-sequence''. A small number of bulge-galaxies are undergoing a
``starburst'' i.e. there has been a sharp increase in the galaxy's
star formation rate over a timescale that is short ($\sim10^8$ years)
in comparison to the age of the galaxy. They differ from blue-sequence
galaxies only in the strength and/or short timescales of the star
formation episode. These galaxies are identified by their unusually
weak Balmer absorption lines, strong UV-blue continua, and weak
4000\AA\ breaks i.e. spectra dominated by light from O/B stars. These
objects lie in the lower left of Fig. \ref{fig:pca}.  As the starburst
ages, the Balmer absorption lines increase in strength as the galaxy
passes into the post-starburst phase \citep{Dressler:1983p2486,
  Couch:1987p2487}, i.e. A/F star light dominates the integrated
galaxy spectrum. These objects lie in the upper left of the figure.
The shape of the left hand side of the distribution in
Fig. \ref{fig:pca} therefore describes the evolutionary track of a
starburst galaxy, with time since the starburst increasing from bottom
to top, and burst strength increasing from right to left
\citep{Couch:1987p2487, 2003MNRAS.346.1055K, wild_psb}. Thus the
evolution of the strongest starbursts that occur in bulges in the
local Universe defines the outer lefthand edge of the distribution.

The unique advantage of our sample for timing the starburst-AGN
connection is that the time since the onset of the starburst can be
measured accurately using population synthesis models. For
ordinary blue-sequence galaxies, the time since a recent, smaller
change in star formation rate is essentially unconstrained using integrated
spectra alone, making a similar measurement of the relative timing of
star formation and black hole growth impossible with the data in hand. 

To sample the properties of AGN in different evolutionary stages of
the starburst, we select a constant number of the strongest starburst
galaxies per unit age. We cannot use our indices to measure the age of
starbursts to arbitrarily old ages, firstly because the light from
weaker starbursts fades rapidly until they become indistinguishable
from ordinary starforming galaxies, and secondly because even the
stronger starburst galaxies suffer from a ``burst age-mass
degeneracy'' at late times. We therefore trade maximum burst age
against sample statistics in order to select a complete sample of 400
bulge-galaxies which have undergone a starburst in the last
0.6\,Gyr. The starburst sample is shown as coloured dots in
Fig. \ref{fig:pca}. 

In the following subsections we give further details of the
sample selection and the calculation of the age of the starburst.

\subsection{Details of the sample selection}

Our bulge-galaxy sample contains galaxies with stellar surface mass
density $\mu^*=0.5{\rm M^*}/\pi \times{\rm r}^2 > 3\times10^8{\rm
  M}_\odot/{\rm kpc}^2$, where stellar mass ($\rm M^*$) is measured
from the 5-band SDSS photometry (J.~Brinchmann,
http://www.mpa-garching.mpg.de/SDSS) and radius (r) is calculated from
the $z$- band Petrosian half light radius.  Although this selection
does not necessarily isolate solely those galaxies with large
bulge-to-disk mass ratios, this mass density cut is based on previous
results that find that AGN in SDSS are predominantly located in host
galaxies with high stellar surface mass density
\citep{2003MNRAS.346.1055K,heckman04}.


We note that our stellar surface mass density cut excludes most,
although not all, pseudo-bulges. Recently, Gadotti \& Kauffmann
(2009)\nocite{Gadotti:2009p2510} have shown that pseudo-bulges contain
4\% of the black hole mass in the local Universe. Empirically, we find
that SDSS galaxies with pseudo-bulges have a slightly lower limit on
their stellar surface mass density of $\mu^*>1\times10^8{\rm
  M_\odot}/{\rm kpc}^2$ \citep{Gadotti:2009p2216} and therefore
we have verified that our conclusions remain unchanged when we use
this lower stellar surface mass density limit.

We impose a per-pixel signal-to-noise ratio (SNR) limit on the spectra
of 8 in the $g$-band to ensure accurate measurement of the stellar
continuum and emission lines. At these low redshifts and relatively
high stellar masses of bulge-galaxies, only ~4\% of spectra have SNR
below this cut and decreasing the limit from 8 to 6 does not alter our
conclusions.

We remove galaxies with stellar velocity dispersions $<$70\,km/s, below which
the SDSS spectra have insufficient spectral resolution to measure
accurate stellar velocity dispersions and therefore reliable black
hole masses. This effectively places a lower limit on black hole mass
of $10^{6.3}{\rm M_\odot}$ (Tremaine et al. 2002), and we emphasise that our results directly
pertain only to black holes with masses above this limit. As with any
hard cut on sample properties, this cut on stellar velocity dispersion
may introduce a small bias into our results: new systems will enter
the sample at later times as the black holes grow and cross the sample
selection threshold. However, with typical black hole growth factors
of 10\% (see Section \ref{sec:disc}) the effect should be small.

Finally, we remove a very small number (five) of candidate ``dusty
starbursts'' which have very large Balmer decrements ($>$2.5 times the
intrinsic ratio, see below) and are identified as having H$\alpha$
luminosities that are too high for their starburst age.  Due to the
obscuration of the youngest and hottest stars behind optically think
dust clouds, these dusty but actively star forming galaxies may
potentially have continuum spectral indices which place them,
incorrectly into the strong-starburst track \citep[][Wild et al. in
prep]{2000ApJ...529..157P}. The inclusion or not of these 5 objects
does not alter the conclusions of this paper.

\begin{figure}
\includegraphics[width=89mm]{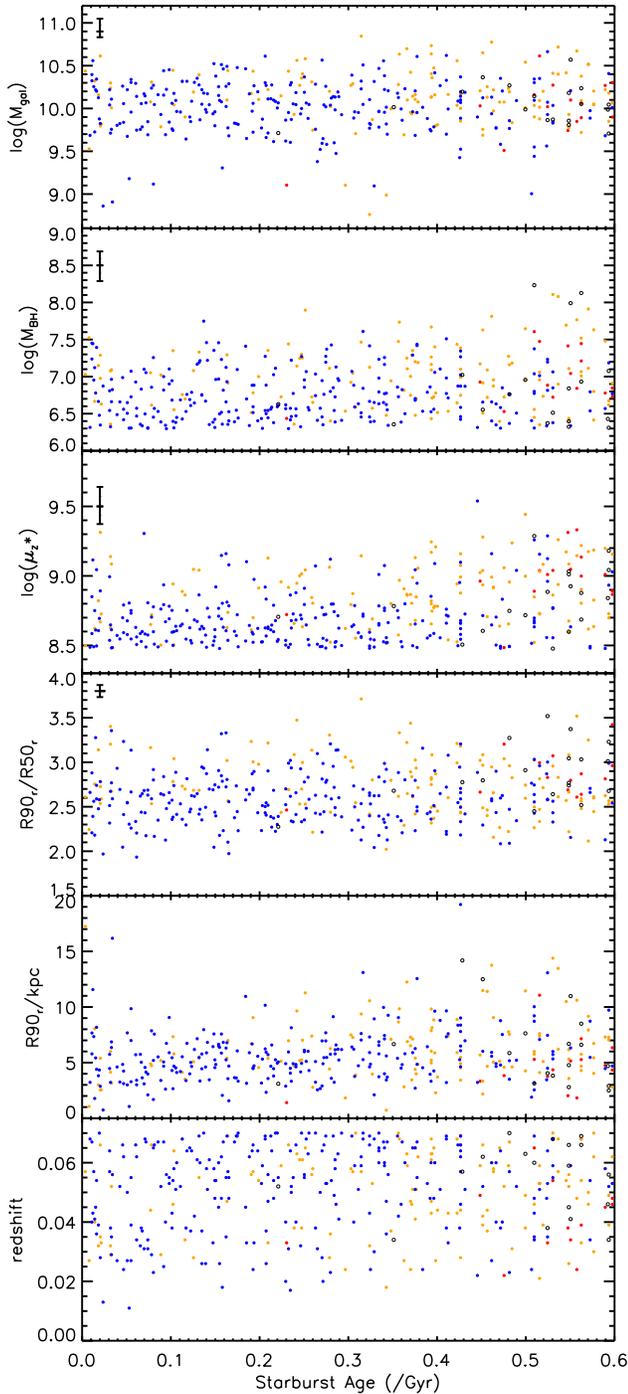}
\caption{Trends of key galaxy properties with starburst age. In all
  panels, blue points indicate star-forming galaxies, orange indicate
  starforming-AGN composites, red indicate pure AGN and black open
  circles are unclassified due to having too weak lines. Typical
  errors are noted in the upper left of each panel where
  significant. From top to bottom as a function of time since the
  onset of the starburst: (i) total galaxy stellar masses, calculated
  from the photometry and mass-to-light ratio taken from the
  likelihood distribution of the starburst models described in Section
  \ref{sec:models}; (ii) black hole masses, measured from stellar
  velocity dispersion and assuming the M-$\sigma$ relation
  \citep{2002ApJ...574..740T}; (iii) stellar mass surface density,
  measured from the total stellar mass and the Petrosian half light 
 radius in the $z$-band; (iv) concentration index, measured from the
  $r$-band Petrosian radii which enclose 50 and 90\% of the light; (v)
  physical size, measured from the $r$-band Petrosian radius which
  encloses 90\% of the light; (vi) redshift. } \label{fig:trends}
\end{figure}

\subsection{Measuring starburst age}\label{sec:models}

In order to boost the signal-to-noise ratio of traditional indicators
of recent star formation history, in particular the measurement of the
Balmer absorption lines, we use indices defined from a Principal
Component Analysis (PCA) of the 4000\AA\ break region of the spectra.
The full method is described in detail in Wild et
al. (2007)\nocite{wild_psb}.  To measure the age of the starbursts, we
build a model library of $10^7$ star formation histories composed of
an old bulge population with superposed exponentially decaying
starbursts \citep[][Charlot \& Bruzual, in
prep]{2003MNRAS.344.1000B}. This star formation prescription, in
addition to being intuitive for the sample at hand, provides a model
library that fully covers the observed distribution of index values
and naturally explains the observed shape of the distribution in
Fig. \ref{fig:pca}.

The bulge population is formed from a tophat starburst of 0.5\,Gyr
width, evolved passively for 13\,Gyr. The star formation history of
the model bulge population is unimportant, but the age is set to give
a precise match to the position of the red-sequence in our stellar
continuum indices.  The metallicity of all models in the library is
fixed to 0.5\,$\rm Z_\odot$ to agree with the position of the peak of
the red-sequence for our galaxy sample. We note that this is slightly
lower than generally measured for samples of SDSS galaxies with masses
$>10^{10}{\rm M_\odot}$ \citep{Panter:2008p2182,
  Gallazzi:2006p2206}. Each method uses different stellar continuum
wavelength ranges and features, and thus the discrepancy may reflect a
real difference in the metallicity of stars contributing most light to
the bluer wavelengths, or a small inconsistency in the stellar
population synthesis modelling. Bursts of star formation with
exponentially declining star formation rates (SFR$\propto \exp^{(-{\rm
    t}/\tau)}$) are superposed on this old stellar population,
allowing burst mass fraction, burst age, decay rate and dust to vary
with uniform priors. Burst age is allowed to range from 0 to 1.5\,Gyr,
burst mass fraction from 0 to 50\% and dust ($\tau_V$) from 0 to
2. While the decline in star formation subsequent to a starburst is
not constrained from the integrated spectrum of an individual galaxy,
the entire distribution of continuum spectral indices observed in
galaxies does provide some additional constraint. For a model in which
star formation rate declines exponentially, $\tau$ values much greater
than 0.3\,Gyr are excluded as they would not produce the observed
population of post-starburst galaxies \citep{Wild:2009p2609}. Very
short values of $\tau$ are equally excluded, because a burst that is
strong enough to reproduce the starburst galaxies with short $\tau$,
evolves along a track which lies to the left of the observed edge of
the distribution in Fig. \ref{fig:pca}, where no galaxies are
observed.  Requiring consistency between the decay in \ha\
  luminosities, which probe instantaneous star formation rates, and
  the age estimated from the continuum provides an additional
  constraint.  Finally, we allow exponentially declining starbursts
  with $200<\tau/{\rm Myr}<350$. We note that this starburst timescale
  is considerably longer than the ``instantaneous'' starbursts
  often assumed in the literature when measuring the ages of
  stellar populations in starburst and post-starburst galaxies. Our
  longer decay timescales result in slightly older derived ages, and
  may explain the population of galaxies with both strong Balmer
  absorption and emission lines (so-called e(a) galaxies)
  \footnote{More complicated star formation histories, such as those
    with an initial slow decline and fast later decline, could also
    produce the observed population of post-starburst
    galaxies. However, this would only result in an small overall
    stretching/contracting of the starburst age axis in places. }.

For each galaxy, the age of the starburst is estimated from the mode
of the probability distribution function of model ages in the standard
way. Starburst ages are well constrained, distributions are unimodal
until beyond the maximum age presented in this paper, and ages are not
sensitive to priors or dust reddening effects of the stellar
continuum.  Precise starburst mass fractions for individual objects
are less well constrained than starburst ages. Rather than relying on
estimated burst mass fractions to define our sample of the strongest
starburst galaxies, we select a constant number of galaxies with the
lowest 4000\AA\ break strength at each starburst age. As described
above, the total number of starbursts selected for the
strong-starburst track sample trades completeness at late ages against
sample statistics. At late ages, old/strong starbursts become
degenerate with younger/weaker starbursts.  Selecting 20 starbursts
per 30\,Myr time bins gives a statistically adequate total sample of
400 galaxies complete to 600\,Myr without risking the inclusion of
galaxies at later times which are in fact younger, weaker
starbursts. 600\,Myr is long compared to the expected timescale of AGN
accretion events ($10^7$ to $10^8$ years), and our results present
``volume-averaged'' quantities, averaging over a large enough sample
of galaxies to smooth out the stochastic nature of AGN activity.

\subsection{Summary of sample properties}
To summarise our selection criteria:
\begin{itemize}
\item Our base catalog is drawn from the SDSS DR7 spectroscopic
  catalog of spectroscopically identified galaxies with extinction
  corrected $r$-band Petrosian magnitude $14.5<r<17.7$.
\item We isolate galaxies with bulges based on their stellar surface
  mass densities  $\mu^*>3\times10^8{\rm M_\odot}/{\rm kpc}^2$.
\item Our sample is limited to $0.01<z<0.07$ to ensure the spectrum
  observed through the 3\arcsec\  fibre aperture is dominated by the
  light from the bulge.
\item Due to the velocity resolution of SDSS spectra, we only accept
  galaxies with stellar velocity dispersion, measured within the
  fibre, of $>70$km/s.
\item From this base catalog, we select the 400 bulges undergoing the
  strongest central starbursts, with starburst age $<$600\,Myr. This
  sample is complete in the sense that we select an equal number of
  starbursts per unit age.
\end{itemize}

The starburst sample is shown as coloured dots in Fig. \ref{fig:pca},
the smoothness of the line traced by its right-hand edge is impressive
and supports our interpretation of the distribution of the spectral
indices.  The stellar masses, black hole masses, stellar mass densities,
concentration indices, sizes and redshifts of galaxies in our
starburst sample are shown in Fig. \ref{fig:trends} as a function of
starburst age. In Appendix \ref{app:images} we present two panels of
images of the youngest and oldest starbursts in our sample. Note in
particular the size of the SDSS fibre through which the spectrum is
taken indicated in the top left of the Fig. \ref{fig:images1}.

The total stellar masses of the galaxies presented in
Fig. \ref{fig:trends} range typically between $5\times10^9$ and
$5\times10^{10}$M$_\odot$. These masses have been estimated from the
probability distribution function of mass-to-light ratios output from
the models described above, and are consistent with those determined
independently from SDSS fibre magnitudes. Based on these model fits,
our sample of starburst galaxies have converted a mass of gas into
stars that is typically around 10-15 percent of the total stellar mass
of stars in the fiber aperture. Because starburst mass fractions
derived from stellar continua are slightly prior dependent, we also
estimate burst masses from the H$\alpha$ luminosities of the galaxies
(presented in the following section). These suggest a slightly higher
typical burst mass fraction of 25 percent.

\section{Black hole accretion rate}\label{sec:3}

We are interested in two quantities that describe the growth of the
black hole. Firstly, the black hole accretion rate (BHAR, $d{\rm
  M_{acc}}/d{\rm t}$) that we estimate directly from the extinction
corrected \oiii\ line luminosity using a bolometric correction of 600,
which corresponds to the conversion ${\rm BHAR}/({\rm
  M_\odot/yr})=4\times10^{-10} {\rm L_{OIII}/L_\odot}$. Secondly, the
black hole growth rate ($d{\rm M_{acc}}/d{\rm t}/{\rm M_{BH}}$) which
characterises the timescale for the black hole to double its mass, and
which we quote in units of the Eddington Luminosity\footnote{${\rm
    L_{Edd}/L_\odot}=3.3\times10^4 {\rm M_{BH}}/{\rm M_\odot}$ is the
  luminosity at which radiation pressure due to electron scattering
  balances the inward pull of gravity for optically thin spherical
  accretion onto the black hole. } (${\rm L_{Edd}}$), assuming a
radiative efficiency of accretion of 10\%. The distinction between
BHAR as measured by \loiii, and black hole growth rate as measured by
\loiii/\ledd\ can be confusing. Throughout the paper we refer to these
quantities as ``accretion rate'' and ``growth rate'' respectively.

In order to measure ``growth rates'' we estimate the black hole masses
from the stellar velocity dispersion of the galaxies ($\sigma$) by
assuming the empirically calibrated ${\rm M_{BH}}-\sigma$ relation
\citep{2002ApJ...574..740T}. We note that elliptical galaxies,
classical bulges (bulges in galaxies with disks) and pseudo-bulges
follow slightly different ${\rm \sigma - M_{bulge}}$ relations, and
therefore ${\rm M_{BH}}$ estimated from a ${\rm M_{BH}}-\sigma$
relation will be different from that estimated from a ${\rm
  M_{BH}-M_{bulge}}$ relation
\citep{Hu:2008p3720,Gadotti:2009p2510,Graham:2009p3700}. The offset
can be attributed primarily to the presence of bars. However, as
  our results are based upon an average over a large sample of AGN, we
  do not believe this will introduce large biases. It is possible that
  bars are more prevalent in the youngest starbursts, in which case
  this effect will lead to slightly enhanced black hole mass
  estimates. We will note the implications of this effect where
  necessary in the results and discussions sections.

We separate our starburst galaxies into those with and without
obscured AGN using optical emission line ratios
\citep{1981PASP...93....5B,2003MNRAS.346.1055K,
  2006MNRAS.372..961K}. For galaxies with a contribution to their
emission lines from both star formation and AGN (``composite AGN'' -
the majority of AGN in our sample) we have removed the contribution to
\oiii\ from star formation using an empirically calibrated mixing
algorithm.  In the following subsections we give further details of
these procedures. Firstly, we justify the use of \oiii\ as an
indicator of black hole accretion rate.


\subsection{\oiii\ as a black hole accretion rate indicator}\label{sec:oiii}

For obscured (Type 2) AGN observed in the optical wavelength range the
high ionisation \oiii$\lambda5007$ emission line is the best available
measure of AGN power. The luminosity of
this line, which originates in the Narrow Line Region, has been shown to
correlate with numerous indicators of total AGN power in unobscured
(Type 1) samples, where the central engine can be viewed directly to
obtain bolometric luminosities \citep[e.g.][]{Mulchaey:1994p3535}. For
obscured (Type 2) AGN, \oiii\ luminosity correlates well with hard
X-ray luminosity for Compton thin AGN, with a scatter of 0.51dex over
4 orders of magnitude in luminosity
\citep[][]{Heckman:2005p3473}. Furthermore, because the AGN Narrow
Line Region lies 100's of parsecs from the central engine, \oiii\ is
not affected by dust obscuration from the torus, unlike X-ray emission
which can be heavily absorbed (Compton thick AGN).

The conversion of \oiii\ luminosity into a bolometric AGN luminosity,
from which we estimate black hole accretion rates, is based on the
strong linear correlation between narrow \oiii\ emission and optical
continuum luminosity in unobscured AGN
\citep{Zakamska:2003p3501}. The conversion is discussed in detail in
\citet{heckman04} and \citet{Kauffmann:2009p2608}. However, we note here
the implicit assumptions in this conversion for completeness. Firstly,
the conversion is necessarily calibrated on unobscured AGN. By
assuming this calibration holds for obscured AGN, we assume the AGN
standard model in which \oiii\ is isotropically emitted and the
difference between obscured and unobscured AGN is in the orientation
of the dusty torus relative to our sightline. Secondly, SDSS AGN
samples extend to lower luminosities than the samples used for the
calibration. However, as shown by \citet{heckman04} and discussed
below, the total \oiii\ luminosity of a volume limited sample of AGN
is dominated by AGN with high \oiii\ luminosities.  Removing galaxies
with \oiii\ luminosities below the limit to which the conversion has
been verified has little effect on volume averaged total black hole
accretion rates. Finally, it should be noted that at some level
\loiii\ should depend on the availability of gas clouds to ionise, and
may therefore correlate with host galaxy properties. This may
plausibly account for some of the scatter in the observed relations
between \loiii\ and other indicators of AGN strength. We do not
believe it is likely that there is a substantial increase in the
number of narrow line clouds over timescales of several hundred Myrs,
which could mimic the results found in this paper, especially given
the relatively slow decline in star formation which presumably
originates from the same gas supply.

 
\subsection{Missing lower luminosity and unobscured AGN}

There are two main ways in which the SDSS narrow line AGN sample is an
incomplete census of the local AGN population: (1) loss of low
luminosity/low growth rate AGN when the lines become indistinguishable from nebular
emission from \hii\ regions, or against the background stellar light;
(2) broad line (unobscured) AGN are not included, because the bright
nuclear light would preclude studies of the integrated galaxy stellar
light. 

Obscured, low luminosity AGN constitute the majority of the low-z AGN
population \citep{2005AJ....129.1795H}, and recent deep X-ray surveys
of local galaxies are revealing increasingly high fractions of low
accretion rate AGN \citep{Gallo:2008p3544}. Hard X-ray observations
are the most sensitive probe for (Compton thin) obscured AGN, and such
low accretion rate systems would not be detected against the
background stellar light in SDSS spectra. But, although these ultra low
luminosity AGN are numerous, as noted above their contribution to the
volume averaged growth of black holes is small \citep{heckman04}.

The volume averaged growth of black holes is dominated by high growth rate (L/\Ledd) systems 
\citep[see discussion
in][]{2005AJ....129.1795H}. More specifically, Heckman et al. (2004) find that
the low-$z$ volume averaged black hole growth rate peaks for black
holes with masses of $\sim10^{7.5}$\Msol\ (their Fig. 1) and is
dominated by systems with L/\Ledd$\gtrsim$ a few percent (their
Fig. 3). Therefore, while missing low-luminosity obscured AGN will
have a negligible effect on estimated global BHARs, we may
legitimately be concerned about missing high-luminosity unobscured
AGN. Although the standard model for AGN suggests we can simply scale
our results by the observed ratio of unobscured to obscured AGN,
deviations from this model may affect our conclusions.
We will return to this point where necessary in the discussion.

\subsection{Emission line analysis}\label{sec:mixing}

\begin{figure}
\includegraphics[width=89mm]{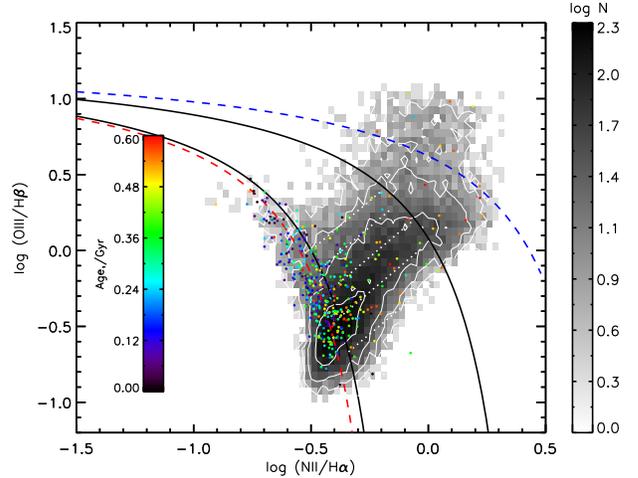}
\caption{The BPT diagram. Bulge-galaxies with all four emission lines
  measured at $>3\sigma$ are shown as gray-scale. The coloured circles
  are the starburst sample, with colour indicating starburst age as in
  Fig. 1. The lines indicate the positions of the demarcation lines
  $C_{\rm SF}$ (dashed red), $C_{\rm AGN}$ (dashed blue),
  $C_{SF/comp}$ (lower black) and $C_{comp/AGN}$ (upper black) as
  described in the text. The white dotted lines show how we assign
  $\kappa$ to composite-AGN when the AGN and starforming contributions
  to the emission lines are calculated (Eqn. \ref{eqn:kappa}).  }\label{fig:bpt}
\end{figure}

\begin{figure}
\includegraphics[width=89mm]{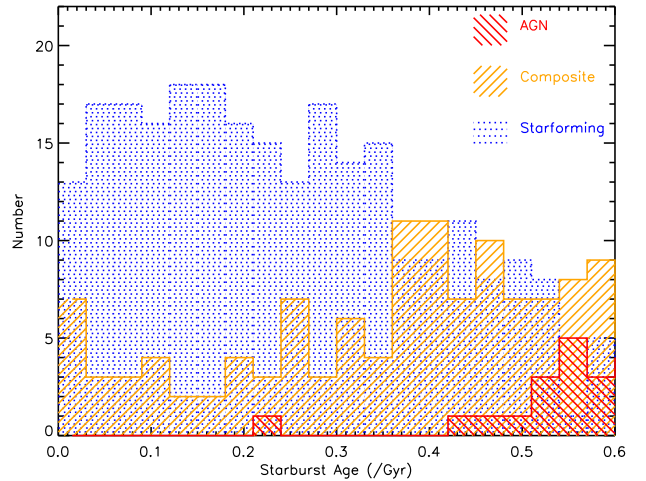}
\caption{An important consequence of our results for the study of the
  AGN-starburst connection in individual objects using wide aperture
  integrated optical spectra alone, is that ``pure'' AGN will not be
  found in starburst galaxies, but only appear after the
  starburst. Before this time the starformation will cause AGN to
  manifest as composite galaxies.  Here we show the number of
  starforming (blue), composite-AGN (orange) and pure-AGN (red) in our
  sample as a function of time since the onset of the
  starburst.} \label{fig:dbpt}
\end{figure}

\begin{figure}
\includegraphics[width=89mm]{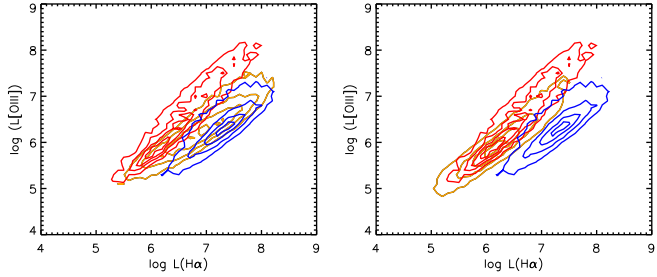}
\caption{For composite-AGN we must separate the
contributions to \oiii\ and H$\alpha$ emission from AGN and star
formation. On the left is the distribution of total \oiii\
vs. H$\alpha$ luminosities for AGN (red), star forming galaxies (blue) and
composite-AGN (orange). From outside to inside the contours encompass
90, 60, 30 and 10\% of the data points. On the right, \oiii\ and
H$\alpha$ luminosities for the composite-AGN have had the contribution
from star formation removed as described in the text. We see that this
correction procedure places the composite-AGN in the same range of
\oiii\ luminosities as the pure AGN.}\label{fig:haoiii}
\end{figure}
 
\begin{figure}
\includegraphics[width=89mm]{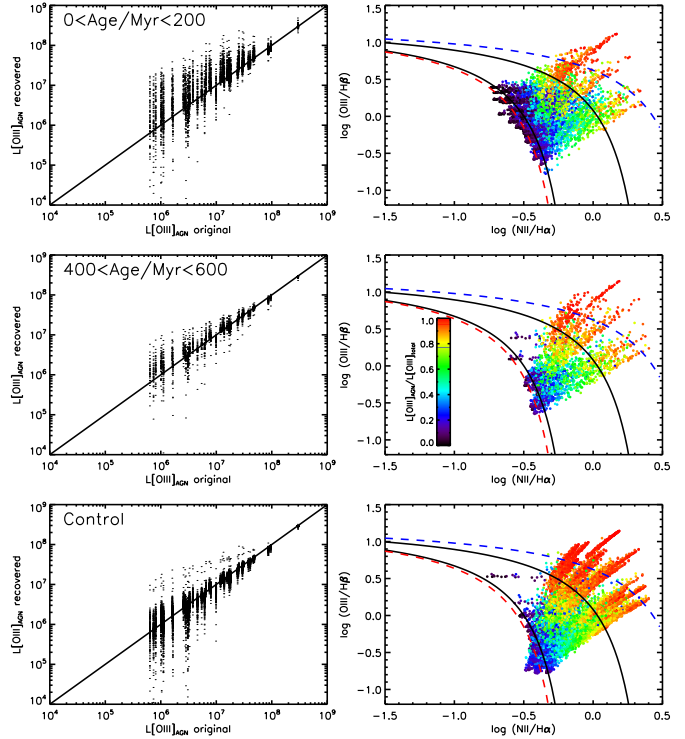}
\caption{In the left panels we show the AGN
contribution to \loiii\ recovered from our Monte Carlo
simulations combining AGN and star-formation-dominated galaxies using the
correction method described in the text, compared to the true AGN
\loiii, for two different ranges of starburst age. In the
right panels, we show the position of all the simulated composite-AGN
on the BPT diagram, with the same demarcation lines as shown in
Fig. \ref{fig:bpt}. The points are colour coded by the AGN contribution to
the total \loiii\ of the simulation galaxy.}\label{fig:corroiii}
\end{figure}

Emission line fluxes and errors, corrected for stellar continuum
absorption, are extracted from the SDSS-MPA catalog
(http://www.mpa-garching.mpg.de/SDSS/), see Brinchman et
al. (2004)\nocite{2004ApJ...613..898T} and Tremonti et
al. (2004)\nocite{2004MNRAS.351.1151B} for details of their
measurement. All lines are corrected for dust attenuation using the
observed H$\alpha$ to H$\beta$ ratio and the two component dust
attenuation curve of Charlot \& Fall
(2000)\nocite{2000ApJ...539..718C}, as presented in Wild et
al. (2007)\nocite{wild_psb} and de Cunha et
al. (2008)\nocite{2008MNRAS.388.1595D}. For star forming galaxies and
composite-AGN (where emission from both star formation and an AGN is
present), an intrinsic H$\alpha$/H$\beta$ ratio of 2.87 is assumed;
for pure-AGN we use a slightly higher H$\alpha$/H$\beta = 3.1$
\citep{1989agna.book.....O, 2006MNRAS.372..961K}. A very small
fraction of our starburst galaxies have lines that are too weak to be
measured accurately, about 4\% of the starburst sample with ages
younger than 600\,Myr and 0.5\% younger than 250\,Myr.  These
fractions are too low to cause any bias in our results.

 In Fig. \ref{fig:bpt} we present the \nii$\lambda6585$/H$\alpha$
  vs. \oiii/H$\beta$ line ratio diagram \citep[the BPT diagram,
  ][]{1981PASP...93....5B} for all galaxies in our sample with lines
  measured at $>3\sigma$ confidence. This diagram is commonly used to
  separate purely star forming galaxies from AGN
  \citep[e.g.][]{2001ApJ...556..121K,2003MNRAS.346.1055K}. The
  starburst sample is overplotted, with colour indicating starburst
  age. We note that the trend for the youngest starbursts to have
  lower metallicities evident in this Figure is expected based on
  previous studies and simulations (Kewley, Geller \& Barton 2006;
  Rupke, Kewley \& Barnes 2010).

The upper black line lies close to the theoretical ``maximum starburst''
contribution from Kewley et al. (2001)\nocite{2001ApJ...556..121K},
above which Seyferts (high \oiii/H$\beta$) and LINERS (low
\oiii/H$\beta$) are easily identified as plumes. As shown by Kauffmann
et al. (2003)\nocite{2003MNRAS.346.1055K} the ordinary star forming
galaxies in the SDSS define a tight sequence that lies far below the
theoretical model maximum, close to the red-dashed line indicated in
the Figure. The simplest interpretation of the objects which lie
between these two populations is that both starformation and AGN
contribute to their emission lines, so-called ``composite AGN''. This
is supported by their younger ages as measured from the stellar
continuum.  The black lines indicate our separation between pure star
forming galaxies and composite AGN (lower black line), and composite AGN and
pure AGN (upper black line). The blue-dashed line is placed to cross through
the center of the pure-AGN. The red-dashed line is placed to cross
through the center of the star-forming galaxies

In this work, all demarcation lines are defined using the same
equation for simplicity:
\begin{equation}
{\rm BPT}_y =1.3+ \frac{0.61}{{\rm BPT}_x - C}
\end{equation}
where the four lines in Fig. \ref{fig:bpt} are defined by $C_{\rm
  SF}=-0.08$ (red), $C_{\rm SF/comp}=- 0.03$ (lower black), $C_{\rm
  comp/AGN}=0.5$ (upper black), and $C_{\rm AGN}=0.9$
(blue). We place the star-forming/composite demarcation
line a little lower than that used by Kauffmann et al. for their
sample. We will discuss the implications of the placement of this line
on our results in the next subsection.

Galaxies which lie below $C_{\rm SF/comp}$ have negligible
contribution to their \oiii\ line from an AGN and we classify them as
star forming galaxies, and those that lie above $C_{\rm comp/AGN}$
have negligible contribution from star formation and we classify them
as pure-AGN. Those that lie in between these two lines we classify as
composite AGN. In Fig. \ref{fig:dbpt} we plot the number of each
emission line class vs. the age of the starburst, from which it is
clear that in optical spectra probing a large spatial area pure-AGN
will only be observed about 500\,Myr after the onset of the starburst
due to contamination of the AGN lines by the strong nebula emission
from star formation. Before this time, AGN will necessarily be
classified as ``composite-AGN''.

In this paper, we include both composite and pure AGN in
our measurements of the total BHAR and distributions of $\rm
L/L_{Edd}$. While galaxies which lie below the $C_{\rm SF/comp}$
demarcation line may contain AGN, these will have very low $\rm
L/L_{Edd}$, therefore this line should be thought of as a limit on
$\rm L/L_{Edd}$ of our AGN sample, rather than a firm division between
galaxies with and without an accreting black hole \citep[see also the
detailed discussion in][]{Kauffmann:2009p2608}. We will return to this
point later when we account for the incompleteness of our AGN sample
as a function of $\rm L/L_{Edd}$.

\subsubsection{Correction of \loiii\ for contamination from star
  formation in composite AGN}
In order to measure the AGN luminosity of composite AGN from the
\oiii\ emission line, we must correct for contamination due to
starformation. Likewise, to measure star formation rates from
H$\alpha$ luminosities, we must correct for contamination from the
AGN. Our method is based upon the assumption that as AGN contribution
increases, galaxies move diagonally away from the starforming sequence
approaching the AGN demarcation line (following the white dotted lines
in Fig. \ref{fig:bpt}). 

Inverting Eqn. 1 to obtain $C$ for an individual composite AGN, we
measure its distance from the starforming sequence to be:
\begin{equation}
D_{\rm BPT} = \frac{C - C_{\rm SF} }{C_{\rm AGN} - C_{\rm SF}}
\end{equation} 
which is normalised such that galaxies that lie in the center of the
star-formation-dominated bulges (red dashed line) and in the
center of the AGN-dominated bulges (blue dashed line) have $D_{\rm BPT}=0$ and
$D_{\rm BPT}=1$ respectively.  

Observing that both pure AGN and star forming galaxies in our sample
show very tight relations between H$\alpha$ and \oiii\ luminosities
(Fig. \ref{fig:haoiii}, left panel), and making the assumption that
$D_{\rm BPT}$ defined in this manner is proportional to the fraction
of H$\alpha$ luminosity arising from the AGN, we derive empirically
the following correction for \oiii\ and H$\alpha$ in the composite
AGN:
 \begin{eqnarray} 
L_{\rm OIII,AGN} &=& L_{\rm OIII,Tot} - \kappa(1 - D_{\rm BPT}) L_{\rm
  H\alpha,Tot} \label{eqn:oiii}\\
L_{\rm H\alpha,SF} &=& L_{\rm H\alpha,Tot} - D_{\rm BPT}L_{\rm
  H\alpha,Tot}\label{eqn:ha}
\end{eqnarray}
where $\rm Tot$ indicates the total contribution to the emission
luminosity from both AGN and star formation.   $\kappa$ is the
  \oiii/H$\alpha$ ratio of starforming galaxies at the point on the
  starforming sequence where the composite AGN originated from
  (following the dashed white lines to meet the red dashed line in
  Fig. \ref{fig:bpt})
\begin{eqnarray} \label{eqn:kappa}
\log_{10}(2.87\kappa) &=& BPT_y^{SF} \\
 &=& 1.3 + \frac{0.61}{BPT_x^{SF}-C_{SF}} \\
 &=& BPT_y + m (BPT_x^{SF} -BPT_x)
\end{eqnarray}
which we solve numerically, setting $m=1.3$ and assuming an intrinsic
\ha\ to \hb\ flux ratio of 2.87 \citep{1989agna.book.....O}. The
success of the separation procedure can be seen in the right panel of
Fig. \ref{fig:haoiii} where the corrected \oiii\ to H$\alpha$ line
luminosities of the composite-AGN follow the relation of the AGN.

\subsubsection{Monte Carlo simulations of the \loiii\ correction procedure}

Because the correction of \loiii\ for contamination from star
formation is not yet common practice in the literature, and has not
been thoroughly tested, we present here a simulation to demonstrate
the validity of the technique. We proceed by taking all pure-AGN with
starburst ages $>250$Myr, and add their lines one-by-one into each
pure star forming galaxy in our starburst sample. We do this for
different age bins of the star-formation-dominated starburst galaxies
and repeat the experiment for a control sample of blue-sequence
bulge-galaxies. This control sample is extracted from the region
defined by the dotted box in Fig. \ref{fig:pca} and is matched to the
starburst sample in black hole mass, stellar mass, redshift and
stellar surface mass density. We then recalculate the position of the
galaxies on the BPT diagram, perform our \oiii\ separation analysis
and compare the output AGN component of the total \oiii\ luminosity,
to the input true AGN luminosity. 

The result is shown in Fig. \ref{fig:corroiii} for two starburst
samples with age bins of $0<{\rm age/Myr}<200$ and $400<{\rm
  age/Myr}<600$, and the control sample of blue-sequence
bulge-galaxies. Similar results are obtained for the intermediate age
starburst sample.  The left hand panels show that on average we
  recover the input \loiii\ well. The right hand panels show the
  simulated composite-AGN on the BPT diagram, colour coded by the
  fraction of \loiii\ contributed by the AGN. On average we see that,
  as expected, galaxies close to the SF-composite demarcation line
  (lower black line) have small contributions, whereas those that sit
  close to the composite-AGN demarcation line (upper black line) have
  contributions close to unity.

  It is noticeable that the youngest starbursts suffer from an
  overestimation of \loiii$_{\rm ,AGN}$ by about a factor of 2 on
  average. This results from their lower mean metallicity which, if
  mixed with a LINER ratio AGN, means they do not follow the assumed
  mixing tracks on the BPT diagram. When mixed with Seyfert line
  ratios, however, the method works accurately. We emphasise that the
  method presented here is suitable for the majority of
  bulge-dominated galaxies at low-redshift. However, for
  low-metallicity star-forming regions there is ultimately a
  degeneracy that prevents a unique solution for the fraction of
  \loiii\ originating from the AGN using these line ratios alone. In
  Section \ref{sec:results} we deliberately exclude the very youngest
  starbursts ($<$30Myr) from some parts of our analysis, which reduces
  the impact of this degeneracy as these galaxies have the lowest
  metallicities (see Fig. \ref{fig:bpt} and Section
  \ref{sec:mixing}). We note in the results and discussion sections
  where this may affect our results.

\subsection{AGN Completeness}\label{sec:compl}
\begin{figure}
\includegraphics[width=89mm]{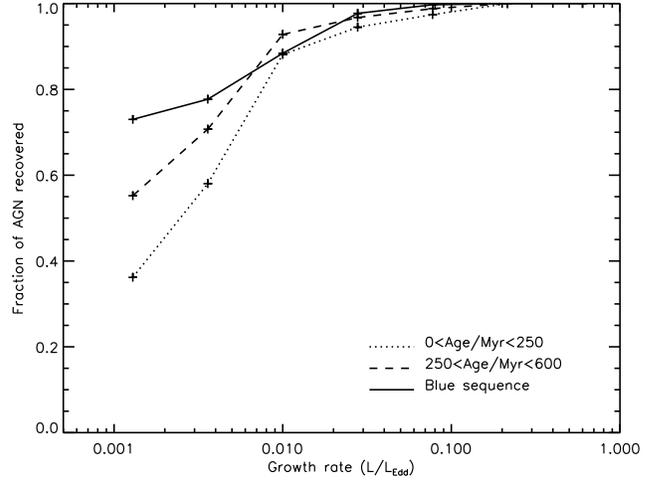}
\caption{Result of the Monte Carlo simulation to
estimate the fraction of AGN recovered as a function of normalised
BHAR ($\rm L/L_{Edd}$) and starburst age.  The dotted and dashed lines
are for the starburst sample with ages as indicated in the figure. The
solid line shows the result for the control sample of ordinary
bulge dominated starforming galaxies. By 250\,Myr after the starburst, the fraction of
AGN recovered is similar to the control sample, however, at younger
ages we can recover fewer AGN due to the masking of AGN lines by the emission
lines from starforming regions. In our analysis we only include
bins which are $>$50\% complete.}\label{fig:lostagn}
\end{figure}

There are three possible causes for not identifying AGN in a purely
optical sample: (1) the AGN emission lines are completely concealed by
the strong nebular lines from star formation; (2) the AGN emission
lines are not identifiable against strong stellar continuum radiation;
(3) the AGN emission lines are obscured by dust.  For the results
presented in this work, we are primarily concerned about correcting
for the first effect, which causes us to preferentially lose weaker
AGN in the youngest starbursts.  Effect (2) will not cause a similar
bias as the red broad-band magnitudes of the sample change little with
the time probed and the low AGN luminosities involved do not
contribute significantly to the time averaged BHAR as discussed in
Section \ref{sec:oiii}. Effect (3) may cause us to loose a small
number of objects at older ages ($>350$Myr) corresponding to the onset
of the Asymptotic Giant Branch phase in lower mass stars, where we
start to see a significant increase in the mean Balmer decrement of
the starburst galaxies. This trend of dust content with starburst age
is the subject of future work.

Returning to the first effect, we use the simulations described above
to calculate the completeness of our samples as a function of ${\rm
  L/L_{Edd}}$ by calculating the position of the simulated composite
AGN on the BPT diagram and finding whether they would be included in
our AGN sample or not (i.e. lie above the lower black line in
Fig. \ref{fig:bpt}). The resulting measured completenesses, shown in
Fig. \ref{fig:lostagn}, are then included in all quoted values of the
AGN fraction. We can see that even at the youngest starburst ages we
recover $>$90\% of AGN with ${\rm L/L_{Edd}}>0.01$, and at later times
this rises to 95\%. Therefore, we are not losing a significant number
of high accretion rate AGN from our samples. However, it is clear that
AGN with ${\rm L/L_{Edd}}<0.003$ are difficult to detect at all ages
during the starburst, with recovery rates of $<$50\%. In order to
investigate further the appearance of low accretion rate AGN,
follow-up observations at higher spatial resolution and, possibly, in
other wavebands will be necessary.

In order to be sure that our key results are completely robust to
changes in the precise positioning of the demarcation between star
forming galaxies and AGN (lower black line in Fig. \ref{fig:bpt}), we repeat
the key results but this time allowing all galaxies which lie above
the red line in Fig. \ref{fig:bpt} to be composite-AGN. We find that
although, of course, there are many more ``AGN'' in our sample these
have low growth rates and the total BHAR remains the same. 

\subsection{Decline in SFR of a starburst}\label{sec:sfrdecline}

\begin{figure}
\includegraphics[width=89mm]{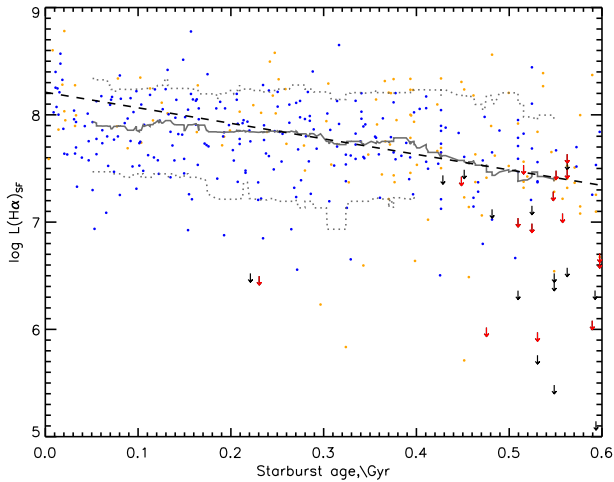}
\caption{H$\alpha$ luminosity from star formation, corrected for dust
  attenuation and contamination from AGN emission as described in the
  text.  Blue points indicate star-forming galaxies, orange indicate
  starforming-AGN composites.  Upper limits indicate unclassified
    objects (black) and pure-AGN (red). The gray solid line and dotted
    gray lines trace the 50th, 10th and 90th percentiles of the
    distribution using a running bin width of 71 galaxies. The black
    dashed line shows an exponentially declining star formation rate
    with $\tau=0.3$\,Gyr, normalised to the median SFR in the first
    40\,Myr. Due to the significant fraction of objects at late ages
    for which we only have upper limits of L(H$_\alpha$) the 10th
    percentile is only indicated to 400\,Myr. } \label{fig:ha}
\end{figure}

Not only does our starburst sample allow us to investigate the time
averaged AGN activity during a starburst, as presented in the results
section below, it also provides a means by which to measure the
form of the decay in star formation after the onset of a starburst. Very
few constraints currently exist on this decay timescale, and
instantaneous or very short starbursts are often assumed in the
literature which are not consistent with the results presented
here. 

In Fig. \ref{fig:ha} we present the \ha\ luminosity of all starforming
and composite galaxies in our starburst sample, as a function of
starburst age derived from the stellar continuum. Composite galaxies
have had the contribution to \ha\ luminosity from AGN removed
(Eqn. \ref{eqn:ha}).  We can clearly identify 3 phases: a short
``starburst'' phase of order 10-30\,Myr, an almost flat ``coasting''
phase which lasts to around 400\,Myr, and a subsequent
decline. Throughout this paper, we have parameterised the overall
decline in starformation by an exponential. In this Figure an
exponential with a timescale of $\tau=0.3$\,Gyr is overplotted as a
black dashed line. It is normalised to the median \ha\ luminosity in
the first 40\,Myr. As discussed in Section \ref{sec:models}, when
fitting starburst ages to our bulge-galaxies we assume a range of
$\tau$ values for the starburst models. This range was decided
  upon by taking the observed decline in \ha\ luminosities into
  account: allowing $\tau$ values much shorter (longer) causes the
  observed \ha\ luminosities of the older starbursts to lie above
  (below) the predictions of the models used to estimate the starburst
  age. While neither stellar continuum nor line luminosities alone
constrain the fall in star formation during a starburst, it is
encouraging that a simple model shows that they can be consistent with
each other. A more precise study of the decline in star formation
following a starburst will be feasible with the data and methods
presented here.

 The starburst timescales of several hundred Myrs measured here
  are consistent with measurements from two entirely independent
  methods. Firstly, from observing starbursts in close galaxy pairs
  \citep{2000ApJ...530..660B,2010arXiv1002.0386F}, and secondly from
  the stellar mass surface density profiles of elliptical galaxies
  \citep{Hopkins:2009p3604}. In the latter paper a very similar form
  for the star formation history is also inferred for the component of
  elliptical galaxies thought to be formed during gas rich
  starbursts.

\section{Results}\label{sec:results}

\begin{figure}
\includegraphics[width=89mm]{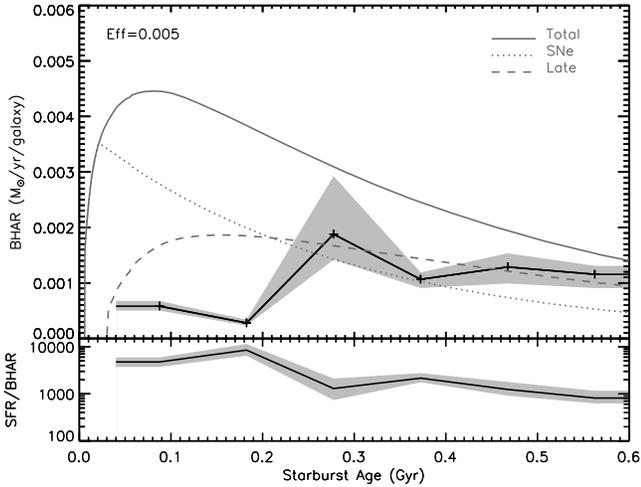}
\caption{Top Panel - the mean BHAR averaged over all galaxies in the
  sample (black line), estimated from a robust mean with 5$\sigma$
  clipping. The gray shaded area shows the typical 10th and 90th
  percentile range on the total BHAR estimated from bootstrap
  resampling of the data. The initial 30\,Myr after the onset of the
  starburst is not plotted, as explained in the text. The model
  prediction for the average gas mass loss rate from the stars formed
  during the starburst (solid gray line) assumes an accretion
  efficiency onto the black hole of 0.5\% of the ejecta from low-mass
  stars between the age of 250 and 600\,Myr.  Stars return mass to the
  interstellar medium through SNe explosions and fast winds from
  massive stars (dotted line) and planetary nebula ejections and
  stellar winds from lower mass stars (dashed line).  Bottom panel -
  the ratio of SFR (from \ha) to BHAR, with mean value and errors
  calculated as for the top panel.}\label{fig:bhar}
\end{figure}

\begin{figure}
\includegraphics[width=89mm]{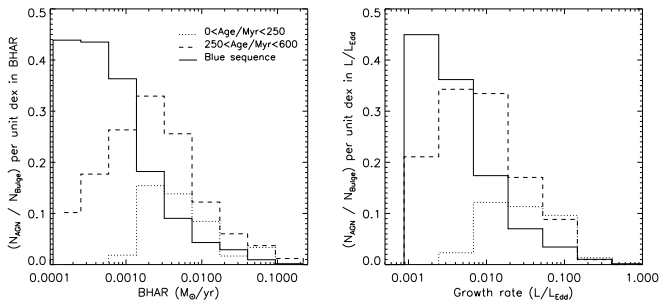}
\caption{The AGN fraction in bulge-galaxies with starburst ages
  $<$250\,Myr (dotted line), with starburst ages between 250 and
  600\,Myr (dashed line) and for bulges undergoing ordinary star
  formation and matched to the starburst sample in stellar mass,
  surface mass density, black hole mass and redshift (solid line, see
  dotted box in Fig. \ref{fig:pca}). {\it Left:} As a function of BHAR. {\it
    Right:} As a function of black hole growth rate (L/\ledd). In both
  panels the AGN fractions have been corrected for incompleteness
  caused by the concealment of AGN emission by emission from star
  forming regions, and only bins which are $>$50\% complete are
  plotted.}\label{fig:accn}
\end{figure}

By averaging over a sample of 400 bulge-galaxies, we can build a
picture of the time-averaged BHAR with respect to the global star
formation activity of the sample, even though individual accretion
events may be relatively short lived. To calculate the average
  black hole accretion rate occurring in each time bin, we sum the
  total \loiii\ originating from the pure- and composite-AGN corrected
  for contamination from star formation, and divide by the total number
  of starburst galaxies in that bin, i.e. including the
  pure starbursts currently not experiencing a strong black hole
  accretion event.  In the top panel of Fig. \ref{fig:bhar} we plot
the mean BHAR (M$_\odot$/yr/galaxy) as a function of time since the onset of the
starburst, finding it to remain low during the early phase of the
starburst, and increase only at ages $>$250\,Myr. A similar offset
between starburst and AGN activity has been suggested for a handful of
nearby galaxies studied in detail
\citep[e.g.][]{Tadhunter:1996p3737,Emonts:2006p1603, Goto:2006p1655,
  Davies:2007p1422}. In the lower panel of Fig. \ref{fig:bhar} we show
how the ratio of SFR/BHAR evolves rapidly with time since the
starburst, where SFR is measured directly from \ha\ luminosity.  We
find that the total mass of stars formed is $2\times10^3$
times the total mass accreted onto the black hole during the first
600\,Myr after the starburst. This ratio is a little higher than, but
of the same order as the ratio of stellar mass to black hole mass in
present-day bulges (Haring \& Rix 2004), suggesting that the processes
we observe in the starbursting bulges may be more broadly relevant to
the co-evolution of bulges and black holes. We will return to this
point in the discussion. We have deliberately not plotted the average
BHAR for starbursts with ages younger than 30\,Myr in
Fig. \ref{fig:bhar} (20 galaxies), because of the additional
incompleteness for these galaxies with extreme star formation rates
and the additional uncertainty in the separation of \oiii\ into
starformation and AGN components (i.e. the initial spike of galaxies
with strong \ha\ fluxes in Fig. \ref{fig:ha}).

To uncover the primary factor driving the low global accretion at
young starburst ages, in Fig. \ref{fig:accn} we plot the distribution
of BHAR (left) and growth rate (right) in the starburst and control
samples. We have divided the starburst sample into two at 250\,Myr and
we compare to galaxies with more ordinary recent star formation
histories (blue-sequence bulge-galaxies matched in black hole mass,
stellar mass, redshift and stellar surface mass density, see dotted
box in Fig. \ref{fig:pca}). All histograms have been corrected for
incompleteness in the samples as a function of L/\ledd\ and starburst
age (Fig. \ref{fig:lostagn} and Section \ref{sec:compl}) and only bins
with an overall completeness of greater than 50\% are plotted.  
  Comparing the distribution of BHAR (left panel) in the three samples
  we see that both young and old starbursts show an enhanced number of
  powerful AGN with respect to the control blue-sequence sample, which
  has been matched to the starburst sample in redshift, stellar mass,
  stellar mass surface density and black hole mass. However, in the
youngest starbursts the AGN fraction is reduced at all accretion rates
relative to the AGN fraction in the older starbursts.  In both young
and old starburst samples the distribution of BHAR turns over at low
BHAR, suggesting that low accretion rate AGN in starbursts are
suppressed compared to in ordinary star forming bulges. However, given
the increasing importance of completeness corrections at lower BHAR
this result should be verified with additional observations.

The right hand panel of Fig. \ref{fig:accn} provides a different view
of black hole accretion patterns during the starburst. Here we see
that the fraction of starbursts with rapidly growing black holes is
enhanced relative to our control sample of blue-sequence
bulge-galaxies. Specifically, the fraction of galaxies with black
holes accreting at rates greater than 0.7\% of the Eddington rate is
$\sim1.2$ times higher in starbursts with ages between 0 and 250\,Myr,
than in ordinary blue-sequence bulge-galaxies, and $\sim2$ times
higher in starbursts with ages between 250 and 600\,Myr. Although
incompleteness prevents us from probing to very low growth rates, the
turn-over in the distribution of L/\ledd\ occurs at higher L/\ledd\ in
the youngest starbursts, compared to both the older starbursts and the
control sample. The shape of these distributions strongly suggest that
low growth-rate accretion events are suppressed during the starburst.

We have not included the 7 composite-AGN with ages younger than
30\,Myr in the dotted histograms of Fig. \ref{fig:accn}, for the
reasons explained above.  The effect of including them is small, and
does not change the impression given by the figures. As discussed in
Section \ref{sec:3}, it is possible that the presence of bars at
certain phases during the starburst may cause biases in the right
panel of Fig. \ref{fig:accn},  due to our estimation of black hole
  mass from stellar velocity dispersion. However, the most likely
  consequence is that we slightly overestimate the black hole mass of
  a small fraction of galaxies with the youngest starburst
  ages. Correcting for this would only enhance our conclusion that the
  highest growth rate events are not suppressed in young starbursts.

Combining all these results we conclude that the global accretion
onto black holes is lower in young starbursts than in older
starbursts. This is caused by an overall drop by around a factor of two in the
number of accretion events at all BHARs above at least
$10^{-3}$M$_\odot$/yr. The possible trend for the turnover in the
distribution of L/\ledd\ to occur at higher values of L/\ledd\ in
starbursts compared to the control sample, and at higher values in
younger compared to older starbursts, should help constrain
the physical mechanism responsible for this decrease. We will return to
this point in the discussion.

\subsection{Summary of observational results} 

The existence of some connection between starbursts and AGN has
  been debated for several decades. Sanders et~al. (1988) argue for an
  evolutionary connection on galaxy wide scales between ULIRGs and
  QSOs. Heckman et~al. (1997) argue for a causal connection on nuclear
  scales (Gonz\'{a}lez Delgado et al. 2001; see Veilleux 2001 for a
  review of the subject)\nocite{Sanders:1988p3844,Heckman:1997p1678,
    GonzalezDelgado:2001p1674,Veilleux:2001p1628}. Here we argue for
its existence for the first time based on a large statistical sample
with a well defined control sample. However, we modify the nature of
connection in several ways.  To summarise the key observational
results, we find that:
\begin{itemize}
\item Galaxies with bulges experiencing a strong starburst are more
  likely to host a rapidly growing and rapidly accreting black hole
  than galaxies with bulges experiencing more normal levels of
  star formation.
\item The total mass of stars formed in the first 600\,Myr after the
  onset of the starburst is $2\times10^3$ times the total mass
  accreted onto the black hole. The black holes increase their mass by
  on-average 5\% during this time.
\item On average, the rate of accretion onto black holes rises steeply
  about 250\,Myr after the onset of a burst of star formation.
\item The lower average BHAR in young starbursts compared to old
  starbursts is caused by a decrease by about a factor of two in the
  number of accretion events at all BHARs above at least $10^3$M$_\odot$/yr.
\item Alternatively, our results can be viewed as an enhancement in
  the global accretion onto black holes during a starburst, relative
  to that occurring in ordinary blue-sequence bulge-galaxies, but
  mainly in bulges with starbursts older than 250\,Myr.
\item On the rare occasions that a black hole accretes during the
  early phase of the starburst, it grows rapidly (has L/\ledd $>$ 0.01). 
\item While higher growth rate events are enhanced compared to
  blue-sequence bulge-galaxies, lower growth rate events instead
  appear suppressed. 
\end{itemize}
 We note that our results apply primarily only to black holes in the
mass range $10^{6.5}-10^{7.5}$M$_\odot$, moderate luminosity AGN
(bolometric luminosities of order $10^{42}-10^{44}$erg/s), and
powerful starbursts which increase the stellar mass in the central
$\lesssim$2kpc radius of the galaxy by at least 10\%.

\section{Discussion}\label{sec:disc}

\subsection{Unobscured AGN}

To begin our discussion, we return to one of the assumptions implicit
in our calculations: to extrapolate our results for obscured (Type 2)
AGN to the growth of all black holes, the simplest approach is to
assume the standard unified model for AGN in which the difference
between unobscured and obscured AGN is in the viewing angle
alone. Under this assumption, the effect of our exclusion of
unobscured (Type 1) AGN will mean that we have undercounted the total
amount of black hole growth in our population of starburst galaxies.

There have been many estimates of the ratio of Type 2 to Type 1 AGN in
the local universe, with values ranging between $\sim$3
\citep[e.g.][]{deGrijp:1992p3789} and $\sim$1 (e.g. Hao et
al. 2005). Thus, we would need to increase the rate of black hole
accretion by factors ranging from 1.3 to 2. There is some dispute
about whether the relative number of Type 1 and Type 2 AGN is an
increasing function of luminosity at low redshift. For example, de
Grijp \& Miley (1992) found no evidence for such a dependence based on
the mid-IR luminosity functions, while Hao et al. (2005) found that
the ratio did increase by a factor of 2-4 over a range of about four
orders of magnitude in \oiii\ luminosity. Thus, we may be
underestimating the relative importance of high accretion rate events
by examining only obscured AGN.

A more interesting possibility would be that the Type 1 and Type 2 AGN
are related to one another in an evolutionary sense (and not
distinguished purely by viewing angle). For example, it is physically
plausible that obscured AGN evolve into unobscured AGN as feedback
associated with the AGN and/or the surrounding starburst clear away
much of the obscuring material (e.g. Sanders et al. 1988). In this
case, by excluding Type 1 AGN we may be selectively undercounting the
relative amount of black hole growth during the late phases of the
starburst compared to the early stages. If so, and if this scenario
were applicable to the moderate luminosity AGN studied in this paper,
then this would only increase the sense of the increase in black hole
growth at these late times.

It is interesting in this regard that \citet{Davies:2007p1422} have
measured the nuclear starburst ages of a sample of mostly unobscured
local AGN, finding a time delay between the onset of the starburst and
the onset of significant black hole accretion. The magnitude of the
delay was similar, although a little shorter than, that found in our
study. As pointed out by \citet{Davies:2007p1422}, the Galactic center
also fits into this picture, where the winds from nearby young OB
stars are not being accreted as efficiently as expected
\citep{Baganoff:2003p3615,Paumard:2006p3638}.

\subsection{Black hole accretion from stellar mass loss}

One popular hypothesis for the physical mechanism
responsible for the feeding of black holes is through the accretion of
matter ejected during stellar evolutionary processes
\citep{Norman:1988p1447}. We can investigate the link between
starburst activity and BHAR by examining the mass returned to the
interstellar medium (ISM) by supernova (SNe) explosions, planetary
nebula ejections and stellar winds predicted from standard stellar
population synthesis models \citep[][Charlot \& Bruzual, in
prep.]{2003MNRAS.344.1000B}. The solid gray line in
Fig. \ref{fig:bhar} shows the mass loss rate of stars formed during a
starburst, and the dotted and dashed lines separate this into mass
loss from high mass (predominantly SNe) and lower mass
stars\footnote{Mass loss from low mass stars in the pre-existing old
  stellar population in the bulge is not significant during these
  phases of a strong starburst.}. During the initial phase of the
starburst mass loss from SNe explosions and fast winds from O/B stars
dominates the mass loss budget, however as can be seen in
Fig. \ref{fig:bhar}, mass loss from lower mass stars soon becomes
dominant. The model is normalised to match the total mass of gas per
galaxy converted into stars in our starburst sample, which we have
measured from the H$\alpha$ line luminosity, converted into star
formation rate \citep{1998ARA&A..36..189K} and corrected for
contamination from AGN. The star formation rate of the model declines
exponentially with a timescale of 300\,Myr, in agreement with the
observed fall-off in the H$\alpha$ luminosity arising from star
formation in the sample.

The model for stellar mass loss rate has been placed onto the same
scale as the BHAR by assuming that 0.5\% of the mass returned to the
ISM by evolved low and intermediate mass stars within 250 and 600\,Myr
is accreted by the black hole. This fraction is similar to that
predicted by, for example, the model of Ciotti \& Ostriker
(2007)\nocite{Ciotti:2007p1476} and to that inferred in galaxy bulges
with predominantly old stellar populations
\citep{Kauffmann:2009p2608}.  It is immediately evident that the average
BHAR of the sample does not follow the {\it total} mass loss rate from
stellar processes (solid line), in particular during the early period
of rapid mass loss from SNe (dotted line). One key difference between
mass loss from SNe compared to that from lower mass stars is in the
ejecta velocity: several thousand km/s for SNe compared to several
tens of km/s for lower mass stars. Fig. \ref{fig:bhar} suggests that
the fast ejecta from SNe must escape the nuclear region without
interacting with the central black hole. This is consistent with the
model of Norman \& Scoville (1988) and with detailed observations of
nearby AGN (Davies et al. 2007).

If a simple model for the feeding of black holes through mass loss
from {\it slow} stellar ejecta is correct, then an important question
is whether SNe are simply inconsequential for black hole growth, or
whether they cause enough disruption of the ISM to prevent the
accretion of the slower stellar ejecta by the black hole. Two features
of our results suggest that the latter may indeed be the
case. Firstly, from Fig. \ref{fig:bhar} we see that the BHAR does not
track the mass loss from intermediate and low mass stars at the
earliest times between 100 and 200\,Myr after the starburst. During
this time, supernova rates remain high and may provide the energy
required to inhibit the accretion of slower stellar ejecta and any
residual gas. Secondly, in the right panel of Fig. \ref{fig:accn} we
see that the distribution of AGN fraction as a function of black hole
growth rate changes with starburst age, i.e.  while higher growth rate
events are enhanced compared to blue-sequence bulge-galaxies, lower
growth rate events are instead suppressed. This pattern is stronger in
young starbursts than in old starbursts, and the turn-over in the
distribution of L/L$_{\rm Edd}$ occurs at higher values for young
starbursts than for old starbursts. A scenario in which growth rate is
linked to the mass of an accreted cloud, and supernovae preferentially
disrupt/expell the lowest mass clouds, would account for the observed
suppression of low growth rate events during starbursts.

\subsection{Alternative models for black hole accretion}

Our observations have direct implications for theoretical models that
attempt to explain the joint evolution of black holes and galaxies
\citep{Silk:1998p2398, Haehnelt:1998p2406, 2006ApJS..163....1H,
  Ciotti:2007p1476}. Our results clearly show that the rate of growth
of the black hole is offset in time with respect to the {\it total}
rate of stellar mass-loss. Moreover, the ratio of the mass accreted by
the black hole to the mass lost by intermediate mass stars alone rises
as the starburst ages. However, there are no direct observational
constraints on the primary physical mechanism responsible for the
feeding of a black hole. In the previous subsection we have chosen to
focus our comparison on the scenario in which black holes feed
primarily on stellar ejecta from intermediate-mass stars, due to its
simple application to our data through stellar population synthesis
modelling. However, we have found that this model in its simplest form
is not sufficient to explain our results. A plausible additional
component to the model, allowing it to match our results, is that
supernovae disrupt the gas in the nuclear regions and prevent the
early accretion of slower material ejected by lower mass stars. We now
discuss two alternative scenarios which are not currently advanced
enough to provide direct comparison with our data.

One alternative for the origin of the time delay is that stellar
ejecta are initially intercepted by gas in the nucleus not yet
converted into stars. Numerical simulations which could test this
scenario are not yet available.  A second possibility is that the
delay in black hole growth could be purely dynamical in nature,
relating to the collision of two bulges during a merger. Some of the
images in Figs. \ref{fig:images1} and \ref{fig:images2} show galaxies
which have experienced significant recent disruption. In simulations
of major mergers with a black hole in each progenitor galaxy, rapid
accretion occurs after coalescence of these black holes, which lags
behind the peak of the starburst by a short time \citep[$\sim 0.1$Gyr,
e.g. ][]{2005Natur.433..604D,2008arXiv0802.0210J}. Although this delay
appears too short to match our results, higher spatial resolution
simulations should be developed before this possibility is ruled
out. Simulations are needed that better resolve the complex physical
processes occurring in a galactic nucleus during a starburst. Although
both of these alternative scenarios for the feeding of black holes may
plausibly lead to a delay in accretion after a starburst, the
intriguing suppression of low growth rate accretion events during the
starburst may still require the presence of some form of feedback
process. In order to gather evidence for or against each of the
scenarios discussed here, spatially resolved spectroscopy would be
invaluable, allowing us to identify double nuclei, measure
starformation rates as a function of radius, and obtain more accurate
accretion rates.

Kauffmann \& Heckman (2009) found that the distribution in the growth
rates of black holes at low-redshift swaps from a power-law
distribution for black holes in old bulges (little or no
star-formation) to a log-normal distribution for black holes in young
star-forming bulges. They showed that the accretion rates in the old
systems were consistent with the black hole accreting 0.3 to 1\% of
the mass loss from evolved stars in the bulge (consistent with the value we
derive). They speculated that the transition to the log-normal regime
was associated with the black hole regulating its own growth in the
young gas-rich bulges. In this paper, we are suggesting that stars may
not only provide the fuel source, but also the feedback (in the form of
supernovae associated with actively star-forming bulges). 


\subsection{Implications for Starbursts}

Our sample of starburst galaxies has other useful applications,
besides tracing the growth of black holes, such as measuring the decay
in star formation following a starburst (Section
\ref{sec:sfrdecline}).  It is clear both from these observational
results, and from simulations of starbursts induced by mergers that
the assumption of an exponential for the decay in star formation
following a starburst is an oversimplification. Given the apparent
complexities in star formation histories suggested by these
simulations, it is perhaps surprising that our simple approach
produces noticeable trends and easily interpretable results. With the
development of galaxy simulations with higher spatial resolution, the
inclusion of more realistic physical prescriptions of star formation,
and the conversion of output into ``observable'' properties, in the
near future we should be able to compare simulations directly with the
sample presented in this paper to further constrain the true star
formation histories of starbursts.

\subsection{Global implications}

Averaged over the full sample, we find that the black holes increase
their mass by 5\% during the first 600\,Myr after the starburst. While
clearly not the dominant phase of the growth of the black holes, this
value is far from insignificant. We have purposefully restricted our
sample to the strongest starbursts that occur in bulges in the local
Universe, simply to enable an accurate measure of the star formation
timescales. However, inspection of the images in Figs. \ref{fig:images1} and \ref{fig:images2} implies that our sample
is not composed exclusively of rare major mergers, and the same time
delay may equally apply to more ordinary episodes of star
formation. At higher redshift, where gas fractions are higher, and
star formation levels are enhanced, the same patterns observed in the
strongest starburst systems at low redshift are expected to become
more widespread. This would imply an increased fraction of high
accretion rate AGN with redshift and, possibly, a decrease in the
fraction of low accretion rate AGN. Growth rates would similarly be
enhanced.

Over the lifetime of a galaxy the mass loss budget from stellar
evolutionary processes is dominated by intermediate and low mass stars, thus our
results are consistent with a scenario in which the tight observed
correlation between black hole and bulge mass is maintained through
the consumption by the black hole of a fixed fraction of low-velocity
winds expelled by such stars in the bulge. 

To evaluate this quantitatively, we adopt a simple prescription in
which we follow the evolution of the stellar mass loss and black hole
growth over a period of 10\,Gyr (the age of typical present-day
bulges) in a model exponentially declining star formation history with
decline timescale of 300\,Myr. For simplicity, we assume that none of
the supernovae ejecta are accreted by the black hole and that none
(0.5\%) of the low-velocity stellar ejecta are accreted before (after)
a time of 250\,Myr. The late-time accreted fraction we adopt is
consistent with observations of black hole growth in old bulges
\citep{Kauffmann:2009p2608}. This simple model predicts that at a time
of 10\,Gyr, the ratio of the remaining stellar mass to the accumulated
black hole mass will be $\sim400$.  Observations of present day bulges
yield an actual mean ratio of 700
\citep{Marconi:2003p2594,Haring:2004p2585}. Given the uncertainties in
converting our measured \oiii\ luminosities into black hole accretion
rates, uncertainties in the stellar mass-loss rates, and the simple
model we have adopted, the similarity of these two values is quite
intriguing. It suggests that the processes at work in our local
starburst sample may well be relevant to the co-evolution of black
holes and bulges over cosmic time, and that black hole growth has been
fueled predominantly by mass loss from intermediate mass stars.

If stars are the primary source of fuel for growing the black hole and
also play a major role in the feedback processes that limit this
growth, then it starts to be less puzzling that black hole mass is
linked to the stellar mass of the bulge.

\section*{Acknowledgements}
VW gratefully acknowledges support from a Marie Curie Intra-European
fellowship. We would like to thank in particular Dimitri Gadotti,
Crystal Martin and Barnaby Rowe for helpful discussions. We
acknowledge the anonymous referee for their useful comments which
helped improve the manuscript. VW also acknowledges the astronomy
groups at the University of Washington, UCSB, UCSC, Stanford and
Caltech who all contributed useful feedback on this work when
presented prior to publication.

Funding for the SDSS and SDSS-II has been provided by the Alfred
P. Sloan Foundation, the Participating Institutions, the National
Science Foundation, the U.S. Department of Energy, the National
Aeronautics and Space Administration, the Japanese Monbukagakusho, the
Max Planck Society, and the Higher Education Funding Council for
England. The SDSS Web Site is http://www.sdss.org/. The SDSS is
managed by the Astrophysical Research Consortium for the Participating
Institutions. The Participating Institutions are the American Museum
of Natural History, Astrophysical Institute Potsdam, University of
Basel, University of Cambridge, Case Western Reserve University,
University of Chicago, Drexel University, Fermilab, the Institute for
Advanced Study, the Japan Participation Group, Johns Hopkins
University, the Joint Institute for Nuclear Astrophysics, the Kavli
Institute for Particle Astrophysics and Cosmology, the Korean
Scientist Group, the Chinese Academy of Sciences (LAMOST), Los Alamos
National Laboratory, the Max-Planck-Institute for Astronomy (MPIA),
the Max-Planck-Institute for Astrophysics (MPA), New Mexico State
University, Ohio State University, University of Pittsburgh,
University of Portsmouth, Princeton University, the United States
Naval Observatory, and the University of Washington. 



\begin{appendix}

\section{SDSS images of starbursts}\label{app:images}
\begin{figure*}
\includegraphics[width=180mm]{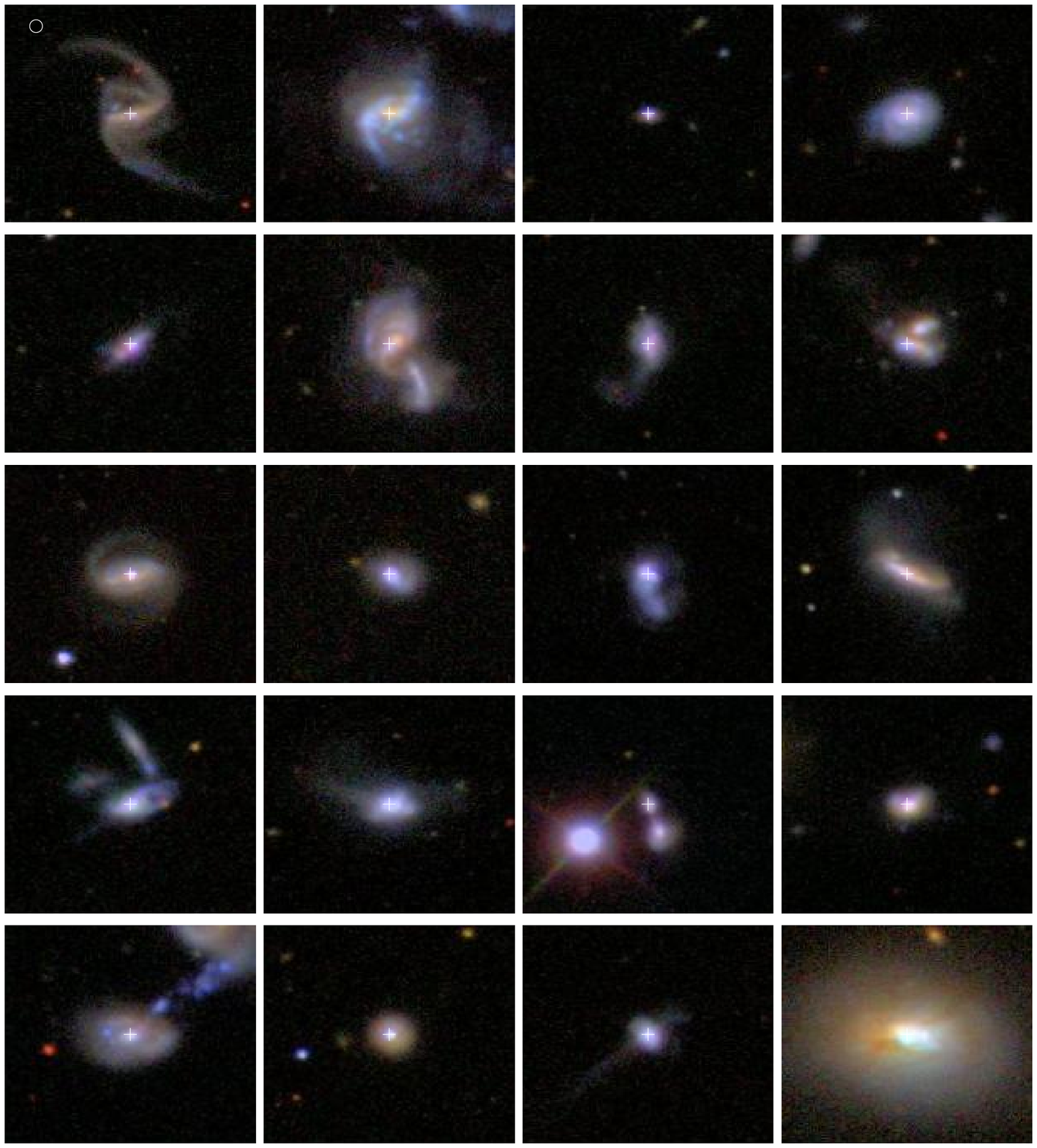}
\caption{SDSS postage stamp images of the 20 youngest starbursts in our
  sample, each is 1\arcmin\ across. In the top left the white circle
  shows the size of the SDSS fibre through which the spectra are
  obtained. The white cross indicates the center of the image and
  approximate position of the fibre. }\label{fig:images1}
\end{figure*}

\begin{figure*}
\includegraphics[width=180mm]{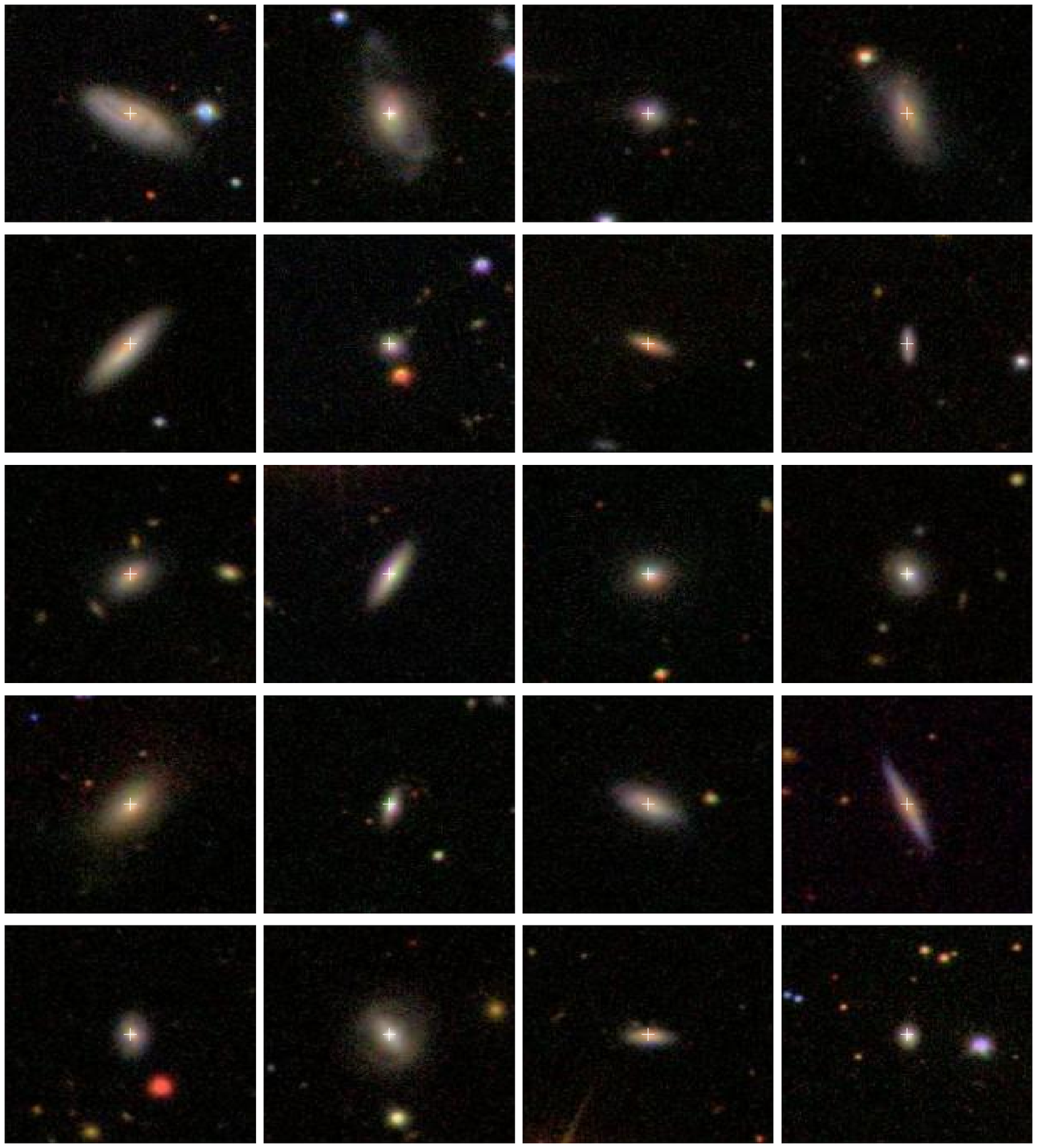}
\caption{1\arcmin\ SDSS postage stamp images of the 20 oldest starbursts in our sample. }\label{fig:images2}
\end{figure*}

\end{appendix}

\end{document}